# A versatile clearing agent for multi modal brain imaging

Irene Costantini[1], Jean-Pierre Ghobril[4], Antonino Paolo Di Giovanna[1], Anna Letizia Allegra Mascaro[1], Ludovico Silvestri[1], Marie Caroline Müllenbroich[1], Leonardo Onofri[1], Valerio Conti[6], Francesco Vanzi[1,7], Leonardo Sacconi[2,1], Renzo Guerrini[6], Henry Markram[4], Giulio Iannello[5], Francesco Saverio Pavone[1,2,3*]

1. European Laboratory for Non-linear Spectroscopy, University of Florence, Via Nello Carrara 1, 50019 Sesto Fiorentino, Italy
2. National Institute of Optics, National Research Council, Largo Fermi 6, 50125 Florence, Italy
3. Department of Physics and Astronomy, University of Florence, Via Sansone 1, 50019 Sesto Fiorentino, Italy
4. Laboratory of Neural Microcircuitry, Brain Mind Institute, EPFL, Station 15, CH-1015 Lausanne, Switzerland
5. Department of Engineering, University Campus Bio-Medico of Rome, Via Alvaro del Portillo 21, 00128 Roma, Italy
6. Pediatric Neurology and Neurogenetics Unit and Laboratories, Department of Neuroscience, Pharmacology and Child Health, A. Meyer Children's Hospital - University of Florence, Viale Pieraccini 24, 50139 Florence, Italy
7. Department of Biology, University of Florence, Via Romana 17, 50125 Florence, Italy

*corresponding author email: francesco.pavone@unifi.it

**ABSTRACT**

Extensive mapping of neuronal connections in the central nervous system requires high-throughput μm-scale imaging of large volumes. In recent years, different approaches have been developed to overcome the limitations due to tissue light scattering. These methods are generally developed to improve the performance of a specific imaging modality, thus limiting comprehensive neuroanatomical exploration by multimodal optical techniques. Here, we introduce a versatile brain clearing agent (2,2'-thiodiethanol; TDE) suitable for various applications and imaging techniques. TDE is cost-efficient, water-soluble and low-viscous and, more importantly, it preserves fluorescence, is compatible with immunostaining and does not cause deformations at sub-cellular level. We demonstrate the effectiveness of this method in different applications: in fixed samples by imaging a whole mouse hippocampus with serial two-photon tomography; in combination with CLARITY by reconstructing an entire mouse brain with light sheet microscopy and in translational research by imaging immunostained human dysplastic brain tissue .

**INTRODUCTION**

The brain is a complex organ with various levels of organization. To understand how it works it is necessary to study the macroscopic organization of each region as well as single cells activities. Neurons from different areas of the brain can communicate with others several millimeters apart, forming a complex network that has yet to be characterized. Mapping all the connections of the entire brain with microscopic resolution is a tremendous challenge in which technological innovation plays a crucial role.

The most common approach, showed by the Allen Mouse Brain Atlas[1] and the Mouse Brain Architecture Project[2], is to slice the tissue into thin sections, then stain and image the sections with various techniques. However, precise cutting, mounting and imaging take a considerable amount of effort and time. Automated, high-throughput imaging methods based on serial sectioning have been recently developed either by embedding the sample in hard resin and slicing it in ultra-thin ribbons which are imaged just after slicing[3,4] or by combining fixed tissue slicing with two-photon microscopy[5]. The major limitations of these techniques are the tissue deformation introduced by slicing and, intrinsically, the destruction of the sample.

An approach that does not require sample cutting is light sheet microscopy (LSM). In LSM the sample is illuminated with a thin sheet of light confined into the focal plane of the detection objective, which collects the fluorescence emission along an axis perpendicular to the illumination plane[6]. This technique drastically reduces the imaging acquisition time and achieves excellent resolution at high penetration depths; however, it requires the sample to be transparent. To reduce scattering and to make the tissue transparent the refractive index has to be homogenized inside the sample. To this end different approaches have been developed. High refractive index organic solvents have been used for optical clearing of entire mouse brains[7-9]. Although these methods guarantee high transparency, protein fluorescence quenching and tissue shrinkage limit their applicability.

Other approaches involving water-based optical clearing agents, such as Sca*l*e[10], SeeDB[11], ClearT[12] and CUBIC[13] were developed to allow improved preservation of fluorescence. Each of them, however, presents other non-negligible limitations such as long incubation times, structural alteration and incompatibility with immunostaining.

The challenge of producing large, transparent and fluorescently labeled volumes has been recently addressed by a new approach called CLARITY[14,15]. This method transforms an intact tissue into a nanoporous, hydrogel-hybridized, lipid-free form. By removing membrane lipid bilayers, CLARITY allows high transparency, whole brain immunolabeling and structural and molecular preservation. This method, however, requires an expensive refractive index matching solution (FocusClear[TM]) whose composition is unknown due to patent protection, limiting its practical applicability. Furthermore, in experiments in which immunostaning or whole organ reconstruction is not required, the complexity of tissue clarification is unnecessary, and an easier clearing approach is highly desirable.

In this work we present a versatile, simple, rapid and inexpensive clearing method based on a water-soluble agent, 2,2'-thiodiethanol (TDE)[16-18]. First, we characterize the effectiveness of TDE in clearing brain tissue. We demonstrate that by applying a solution of 47% TDE directly on paraformaldehyde fixed sample we are able to increase the penetration depth of two-photon imaging while preserving protein fluorescence and avoiding structural deformation. We demonstrate that TDE is a suitable agent to perform serial two-photon tomography (STP) with the 3D reconstruction of the whole hippocampus of a fluorescent transgenic mouse with high resolution and sensitivity. Furthermore, since the refractive index of TDE can be finely adjusted it can also be useful as an optical clearing medium for CLARITY. We demonstrate this novel utilization by imaging the whole mouse brain with LSM and reconstructing specific cell type distributions and vascular structures. Finally, we couple TDE clearing with CLARITY immunohistochemistry and stain a large volume of human dysplastic brain tissue. This clearing approach significantly expands the application of single- and two-photon imaging on large samples and provides a novel and useful method for quantitative morphological analysis of the neuroanatomical structure in mouse and human brain.

**RESULTS**

We propose a novel and versatile method to clear brain tissue suitable to multiple image modalities. This clearing method is based on 2-2' thiodioethanol, a non-viscous, aqueous solvent whose refractive index (RI) can be finely tuned between that of water (1.33) and that of oil (1.52) by mixing it with a water-based solution in varying ratios. TDE was tested on paraformaldehyde (PFA)-fixed mouse brain, CLARITY-processed mouse brain and formalin-fixed human brain samples.

**PFA-fixed brain tissue clearing**

We characterized the effect of TDE on fixed brain slices in terms of transparency and preservation of the sample. Mouse brain sections of 1mm thickness were incubated in solutions of increasing percentage of TDE in phosphate-buffered saline (PBS). TDE diffused rapidly and homogeneously through the tissue and after a few hours the sections became transparent. We measured the transmittance of TDE-clarified samples and compared it with a common water-based optical clearing agent (SeeDB; Fig. 1a,b). Every solution showed a transmittance enhancement with higher wavelength and increasing percentage of TDE up to a percentage of 80% TDE/PBS where the transmittance was comparable with that of SeeDB-cleared samples. To perform two-photon fluorescence microscopy (TPFM) imaging we selected 47% TDE/PBS solution which corresponded to the RI matched by our microscope objective (Zeiss 20X Sca*l*e objective, with n=1.42). The clearing did not cause significant linear deformation (10% shrinkage with 47% TDE/PBS solution; Fig. 1c) or anisotropic distortion (see Supplementary Fig. 1). To quantify the improvement of the imaging depth achievable in cleared samples we measured the contrast value decay as a function of depth in uncleared and cleared slices. The penetration depth with TPFM imaging in cleared tissue was almost four times higher than in samples in PBS (Fig.1d,g). By measuring the fluorescence decay in cleared and uncleared tissue, we found that TDE did not increase bleaching (Fig. 1e) nor did it lead to quenching of protein fluorescence (Fig. 1f). The fluorescence intensity remained constant over time, allowing long-term measurements for up to two months. To characterize the ultrastructure preservation of cleared samples we performed transmission electron microscopy (TEM) imaging. The ultrastructure in TDE-treated samples was well preserved with the exception of occasional and localized swelling of myelin sheaths. Nuclei, axons, dendrites, vesicles and organelles, could be easily distinguished. Mitochondria appeared undamaged and ultrastructural features such as synaptic vesicles and postsynaptic densities were well preserved (Fig. 1h).

**Two-photon serial sectioning of whole hippocampus**

The high transparency obtainable with TDE clearing was exploited to expand the imaging depth of the serial two-photon tomography technique. The enhanced penetration depth afforded by TDE clearing allowed reducing tissue slicing, thus decreasing acquisition time, minimizing cutting artifacts and enabling lossless imaging of the whole sample volume. We reconstructed an entire mouse hippocampus dissected from a fixed adult Thy1-GFP-M mouse brain. We imaged the hippocampus with TPFM using a Zeiss 20X Sca*l*e objective (Fig. 2a). Through this complete tomography, specific anatomical features of the hippocampus such as dentate gyrus and *Cornu Ammonis* areas (Fig. 2b) were easily recognizable. Moreover the high resolution stacks (Fig. 2c) acquired in this tomography could reveal fine anatomical details of the sample such as spines and varicosities (Fig. 2d,e). The high sensitivity of this approach enabled the complete tracing of single neurons through a large volume without interpolation (Fig. 2f,g).

**Whole mouse brain imaging with light sheet microscopy**

After demonstrating the possibility of performing complete reconstructions of large volumes at high resolution, we aimed at getting an expanded view of morphological details over the whole mouse brain. LSM, combined with suitable clearing techniques like CLARITY, has the potential of imaging big volumes of tissue in a short time. By using TDE as clearing agent for the CLARITY protocol, we obtained an inexpensive tool to expand the applicability of LSM. In order to match the RI of FocusClear[TM] (RI=1.45) we selected a 63% TDE/PBS solution. This medium preserved entire mouse brains and made them uniformly transparent (Fig. 3a) with a transmittance comparable to that of FocusClear[TM] (Fig. 3b). The electrophoretic step of the CLARITY protocol causes the tissue to expand. We measured the linear deformation of the tissue caused by the clearing with TDE and with FocusClear[TM]. TDE, similarly to FocusClear[TM], shrank the tissue back, producing a final tissue expansion of 16% (Fig. 3c). We processed with CLARITY and cleared with TDE the brain of an adult PV-cre-tdTomato mouse, in which parvabuminergic neurons were labeled with the tdTomato fluorescent protein. We reconstructed a whole mouse brain with sub-cellular resolution (Fig. 4a,b,c). We could easily dissect the main anatomical features and visualize the distribution of cells bodies and axonal bundles over the whole brain. The preservation of the sample was maintained also in a GAD2-cre-tdTomato mouse brain in which GABAergic interneurons were labeled, showing finer details of neuronal connections (Fig. 4b,d) which are useful for tracing and network analysis. To further test the compatibility of this clearing method with organic dyes, we imaged with LSM a mouse brain labeled with the nuclear cell marker propidium iodide (Fig. 4b,e). The fluorescence of this molecule was retained over the whole brain, providing valuable data for automatic cell body counting algorithms[19]. Finally, we implemented the clearing protocol on whole brain vasculature labeled with FITC-albumin (Fig. 4b,f). This approach revealed even the smaller capillaries, allowing access to large volumetric reconstructions of the mouse vasculature.

**Immunostaining and human brain imaging**

After demonstrating the general applicability of this method to a wide range of samples like transgenic animals and labeling with various dyes, we wanted to show its translational potential on human samples. We therefore investigated the compatibility of TDE clearing with immunohistochemistry (IHC). We showed that it is possible to combine the passive CLARITY (PC) method with antibody staining and TDE clearing of large samples. At first we tested the compatibility of TDE clearing with an Alexa Fluor 594 conjugated anti-GFP antibody staining on a slice of a PFA-fixed Thy1-GFP-M mouse brain. The characteristic features of GFP-expressing neurons (for example dendritic spines) were easily recognizable in the TPF images (Supplementary Fig.2), suggesting that the 47% TDE/PBS solution did not affect the protein-antibody interaction. After proving that TDE is a valid clearing medium for mouse IHC stained tissue, we applied the protocol to human brain by studying a large specimen surgically removed in a patient with drug resistant epilepsy due to hemimegalencephaly (HME). Hemimegalencephaly is a malformation of cortical development in which one cerebral hemisphere is enlarged, exhibiting a grossly abnormal gyral pattern and an abnormally laminated cortex, harbouring abnormal cell types cells (dysmorphic giant neurons and balloon cells)[20]. We chose this severe form of cortical dysplasia as it to ideally suit our purpose of testing the method's sensitivity in highlight different elements of abnormal cytoarchitectonic organization.

A 2-mm thick block of cortex, stored in formalin, was treated with the PC technique, stained with different antibodies and cleared with TDE solution. We were able to label the tissue with antibodies against parvalbumin (PV) (Fig. 5a) and glial fibrillary acidic protein (GFAP) (Fig. 5b) as well as performed double labelling with the combination of them (Fig. 5c).

We imaged a cube of 1 $mm^3$ of stained tissue with TPFM maintaining the same contrast through the whole depth (Fig. 5d,e,f). This optimization of the PC staining protocol allowed augmenting the penetration depth of the antibody by 100%. We could easily recognize single axons in the densely labeled sample and perform tracing of neurons across the entire volume (Fig. 5g,h). Isolated neuronal processes could be clearly distinguished, demonstrating that projections can now be studied by tracing several neurites over a mm-sized volume.

Dysplastic tissue subjected to CLARITY was sampled from a bigger specimen used to perform routine anatomopathological characterization. This routine allowed a direct comparison between conventional hematoxylin/eosin staining and the immunostaining of the cleared sample. We identified giant dysmorphic neurons using both staining techniques, remarking that the features observed in the cleared tissue were directly comparable to those obtained through conventional staining techniques (Supplementary Fig. 3).

**DISCUSSION**

In the last few years the optical dissection of brain architecture has been addressed with the development of different microscopy techniques coupled with various methods for sample preparation. In this respect, several clearing protocols have been recently developed to reduce light scattering during imaging. Each of them has distinctive characteristics which makes it suitable for a specific optical technique while limiting its use for complementary ones. In this work we presented a simple, quick and inexpensive clearing method based on TDE as refractive index matching agent. The characterization of the TDE clearing was based on several criteria: resulting transparency in terms of light transmittance, linear deformation, fluorescence quenching and imaging depth. We showed that the transparency attainable depended on the RI of the solution and, therefore, on the increasing concentration of TDE. Clearing did not substantially change the final volume of the specimen, nor did it lead to linear deformation or anisotropic distortion. The protein fluorescence of the sample remained constant over time during long-term incubations, even up to months, indicating that our clearing protocol does not lead to quenching. TDE diffused very quickly inside the brain tissue, allowing a fast and easy clearing procedure compared with other techniques such as, for example, Sca*l*e or SeeDB[10,11] (see table 1). The imaging depth achievable after the clearing procedure was increased by a factor of four in PFA-fixed samples. Our protocol was shown to be suitable for the study of endogenous fluorescence in transgenic animals since it did not lead to fluorescence bleaching, conversely to organic solvent techniques[7-9], which can only be applied to immuno-labelled samples or in transgenic animals with a high fluorescence expression level. The low-viscosity of TDE allowed for the complete reconstruction of a large brain area using STP. Here we were able to image an entire PFA-fixed Thy1-GFP-M mouse hippocampus cleared by direct incubation in TDE solution. Every individual part of the hippocampus was imaged at high resolution, giving the possibility to resolve spines and varicosities across the whole area. Combining the acquired stacks by means of an automatic 3D-stitching tool allowed us to trace single neuronal processes with high accuracy throughout the hippocampus. This point underlines a particular strength of our technique since SeeDB, in contrast, is incompatible with STP because of its high viscosity which limits the acquisition of 3D volumes to the imaging depth.

The direct incubation of TDE is not suitable for imaging entire organs. While in some approaches limitations due to insufficient clearing are overcome by slicing the sample in 1-1.5 mm thick slices[21], here we pursued optimum transparency in the whole mouse brain by coupling TDE with the CLARITY technique. We found that TDE is a valid alternative to FocusClear$^{TM}$ as refractive index matching solution, considerably lowering the cost of every experiment and making large-volume, high-throughput imaging with LSM affordable. Moreover, the unknown composition of FocusClear$^{TM}$ impaired any effort of the scientific community to further improve its clearing efficiency. For example, the refractive index variation achievable by different TDE/PBS percentages allows an *ad hoc* optimization of the clearing capability for different tissue types and permits integration of other compounds that can potentially improve the optical transparency of the selected specimen. Finally, the versatility of TDE as multipurpose clearing medium can facilitate correlative approaches to overcome the inherent limitations of a single imaging technique and thus enable multi-modality in brain anatomy[22].

The compatibility of TDE clearing with immunostaining was demonstrated on brain tissue from different species, namely on mouse brain and on human brain samples. We were able to homogenously stain and image a cube of 1 mm$^3$ tissue from a hemimegalencephaly patient and to follow neuronal fibers throughout the whole volume. The capability of performing immunochemistry in a large volume of human brain tissue with micrometric resolution and high sensitivity is a crucial step in the direction of human brain connectomics. It also represents an innovative translational tool for finely characterizing macroscopic and microscopic circuit alterations and identifying cells having aberrant morphology, a common finding in the brain of individuals with refractory epilepsy, intellectual disability and autism. Future work will explore the limitations of the technique in regard to maximum sample thickness to find the best compromise between contrast (which is limited by the penetration depth of antibodies) and sample throughput.

In conclusion, compared with other techniques, our TDE protocol covers a wide range of applications. The most intriguing characteristic of our method lies in its versatility; the possibility to choose the most suitable tool for different experiments permits a powerful investigation of neuronal networks in the brain. We believe that, together with advances in microscopy and computational analysis, our TDE protocol can contribute to enhance our understanding of anatomic structure and connectomics of the brain. In the future, the usefulness of TDE may not be limited only to brain neuroanatomy investigations but, as shown by other clearing methods, could also span different areas of research such as entomology[7,23], embryology[9,24] and even medicine with 3D anatomical studies of biopsies from different organs and tissues[14].

**METHODS**

**Transgenic animal model**

We analyzed different lines of transgenic mice; for the sparse labeling of pyramidal neurons with GFP we used the Thy1-GFP-M line[25]. For the visualization of GABAergic interneurons we used GAD2-ires-Cre-tdTomato mice[26] and for imaging the subpopulation of parvalbumin positive neurons we used the PV-Cre-tdTomato line[27]. The experimental protocols involving animals were designed in accordance with the laws of the Italian Ministry of Health. All experimental protocols were approved by the Italian Ministry of Health.

**Human brain specimen collection**

The human brain sample was removed during a surgical procedure for the treatments of drug resistant epilepsy in a child with HME. The sample was obtained after informed consent, according to the guidelines of the Human Research Ethics Committee of the A. Meyer Children's Hospital. Upon collection, the sample was placed in neutral buffered (pH 7.2-7.4) formalin (Diapath, Martinengo, Italy) and stored at room temperature until the clearing process.

**Preparation of fixed mouse brains**

Adult mice (p56) were deeply anesthetized with an intraperitoneal injection of ketamine (90 mg/kg) and xilazine (9mg/kg). They were then transcardially perfused with 100 ml of ice-cold 0.01 M phosphate buffered saline (PBS) solution (pH 7.6), followed by 100 ml of freshly prepared ice-cold paraformaldehyde (PFA) 4% in 0.01 M PBS (pH 7.6). The brain was extracted from the skull and fixed overnight in 20 ml of PFA 4% at 4 °C. Samples were then rinsed three times (30 minutes each) in 20 ml of 0.01M PBS at 4°C. The brains were stored in 20 ml of 0.01M PBS at 4°C. For transmission electron microscopy (TEM) the anesthetized animal was perfused with 10 ml of PBS immediately followed by 300 ml of a mixture of 2.5% glutaraldehyde (Glut) and 2% PFA in 0.01 M phosphate buffer (pH 7.4). After 2 hours, the brain was removed and stored at 4°C in 50 ml of 0.01M PBS.

**Preparation of CLARITY-processed mouse brains**

CLARITY samples were prepared according to the Chung protocol[14]. Adult mice (p56) were anaesthetized with isofluorane and transcardially perfused with 20 ml ice-cold PBS solution (pH 7.6) followed by 20 ml of a mixture of 4% (wt/vol) PFA, 4% (wt/vol) acrylamide, 0.05% (wt/vol) bis-acrylamide, 0.25% (wt/vol) VA044 in PBS. Brains were then extracted and incubated in the same solution at 4°C for 3 days. The samples were then degassed and the temperature was increased to 37 °C to initiate polymerization. The embedded sample was extracted from the gel and washed with clearing solution at 37 °C through gentle shaking. To perform electrophoretic tissue clearing (ETC), hydrogel-embedded brains were placed in a custom-built organ-electrophoresis chamber. Sodium borate buffer (200 mM, pH 8.5) containing 4% (wt/vol) sodium dodecyl sulfate (SDS) was circulated through the chamber and a voltage of 20V was applied across the ETC chamber at 37 °C for several days. After clearing, brains were incubated in $PBST_{0.1}$ (PBS and 0.1% Triton X-100, pH 7.6) at 37 °C for 2 days to remove the SDS. For vasculature staining, a specialized CLARITY perfusion protocol was applied. Mice were anesthetized with pentobarbital (150 mg/kg) and transcardially perfused with PBS and 10ml of CLARITY monomer solution. After this, a third perfusion was performed with 20 ml of monomeric solution containing FITC-albumin at 4mg/ml. During polymerization the fluorescent albumin tightly integrated into the acrylamide mesh and was therefore not eliminated during lipid removal.

**Preparation of passive CLARITY (PC) -processed samples**

Blocks of fixed samples were washed in PBS at 4°C for one day and then incubated in 4% (wt/vol) PFA, 4% (wt/vol) acrylamide, 0.25% (wt/vol) VA044 in PBS at 4 °C for 2 weeks. The samples were degassed and then the temperature was increased to 37 °C to initiate polymerization. The embedded samples were extracted from the gel and incubated in clearing solution (sodium borate buffer 200 mM, pH 8.5) containing 4% (wt/vol SDS) at 37 °C for two weeks while gently shaking. After clearing, samples were incubated in $PBST_{0.1}$ at 37 °C for 1 day to remove the SDS. Before staining, human CLARITY brain samples were manually cut into pieces of approximately 2 mm$^3$ using a scalpel. The PC protocol has been applied to Thy1-GFP-M mouse brain slices and a human brain bioptic sample.

**Optical clearing with TDE**

Murine PFA-fixed samples were cleared with serial incubations in 20 ml of 20% and 47% (vol/vol) 2,2'-thiodiethanol in 0.01M PBS (TDE/PBS), each for either 1 hour at 37°C or for 12 hours at room temperature (RT) while gently shaking. CLARITY-processed murine brain samples were cleared with serial incubations in 50 ml of 30% and 63% (vol/vol) 2,2'-thiodiethanol in 0.01M PBS (TDE/PBS), each for 1 day at 37°C while gently shaking. Human brain samples were cleared with serial incubations in 10 ml of 20% and 47% TDE/PBS for 10 minutes at 37°C while gently shaking.

**Staining of CLARITY-processed samples**

To perform immunostaining, PC processed samples were incubated at RT for 2 days with the primary antibody (dilution, 1:50) in $PBST_{0.1}$ solution, followed by washing at RT for 1 day in $PBST_{0.1}$ solution. The tissue was then incubated with the secondary antibody (dilution, 1:50–1:100) at RT for 2 days in $PBST_{0.1}$ solution, followed by washing at RT for 1 day in $PBST_{0.1}$ solution. We used as primary antibody an anti-PV (parvalbumin) antibody (Abcam, UK, cat. ab11427 or ab64555) and an anti-GFAP (glial fibrillary acidic protein) antibody (Abcam, cat. ab53554) and as secondary antibody an Alexa Fluor® 568 conjugated IgG (Abcam, cat. ab175471 or ab175704) and an Alexa Fluor® 488 conjugated IgG (Abcam, cat. ab150105). After staining, samples were optically cleared with 47% TDE/PBS before imaging by two-photon fluorescence microscopy in two color channels. For whole brain nuclei staining, CLARITY-processed murine samples were incubated at 37°C for 2 days with 1:50 Propidium Iodide (PI, LifeTechnologies, CA, P3566) solution, in $PBST_{0.1}$ followed by washing at 37°C for 1 day in $PBST_{0.1}$ solution. Subsequently they were optically cleared with 63% TDE/PBS before imaging with a light sheet microscope.

**Measurement of light transmittance and linear deformation**

PFA-fixed Thy1-GFP-M mouse brains were embedded in 4% agarose in 0.01 M PBS and cut into 1 mm coronal sections with a vibratome. The agarose surrounding each half-brain slice was removed and the slices were cleared by serial incubations in 20%, 47%, 60%, 80% and 100% (vol/vol) TDE/PBS, each for 1 hour in 20 ml glass vials at 37°C while gently shaking. Clearing with SeeDB was performed following the protocol described by Ke[11]. Slices were cleared by serial incubation in 20 ml of 20%, 40% and 60% (wt/vol) fructose, each for 4-8 hours and incubated in 80% (wt/vol) fructose for 12 hours, 100% (wt/vol) fructose for 12 hours and finally in SeeDB (80.2% wt/wt fructose) for 24 hours while gently rotating at RT. For CLARITY mouse brains, we followed a different protocol for sample preparation. Since the porosity of the final gel makes samples unsuitable for agarose embedding and vibratome cutting, we obtained 2 mm thick coronal slices from a Thy1-GFP-M mouse CLARITY brain using a rat brain slicer (Alto rat brain coronal matrices, CellPoint Scientific, MD). Slices were then cleared by serial incubations in 20 ml of 20%, 47%, 63% (vol/vol) TDE/PBS, each for 1 hour at 37°C while gently shaking or with FocusClear$^{TM}$ (CelExplorer Labs, Taiwan). Light transmittance was determined using a spectrophotometer (Lambda 950 UV/Vis/NIR Perkin Elmer, MA) with uncleared slices in PBS as reference samples. For the evaluation of linear deformation, sample photos were taken on a glass dish filled with PBS or the respective clearing mediums. Transmission images of whole CLARITY mouse brains were taken after 1 day incubation for each solution of TDE/PBS or after 3 days incubation in FocusClear$^{TM}$. Based on top view photos, the area of the samples was determined using ImageJ/Fiji. The linear deformation was quantified by normalizing the area of the cleared brain with the area of the brain in PBS and calculating the square root of that quotient. To characterize possible nonlinear distortion the edges of brain slices or whole brain were manually traced using GIMP ([www.gimp.org](www.gimp.org)), resized using the linear deformation parameter obtained before, and superimposed using different colors.

**Measurement of fluorescence quenching and bleaching**

GFP fluorescence quenching and bleaching evaluation was performed on uncleared and cleared, PFA-fixed samples imaged with the TPFM. Fixed Thy1-GFP-M mouse brain slices of 2 mm thickness were optically cleared with 47% TDE/PBS at 37°C. To measure the effect of quenching, two photon images were acquired at different times and slices were incubated in 47% TDE/PBS at RT between acquisitions. Freshly made TDE solution was used for every measurement. The mean fluorescence intensity of homogeneous regions (100 x100 µm$^2$ region of interest; ROI) for each time point was measured using ImageJ/Fiji. Bleaching was quantified as the temporal decay of the mean fluorescence intensity in a ROI enclosing a dendrite portion (20 x 20 µm$^2$) and the value of a neighboring area without a dendrite was subtracted as background.

**Evaluation of imaging depth**

A brain slice of 2 mm thickness, from an FBV mouse, was incubated in PBST$_{0.5}$ (PBS and 0.5% triton X-100, pH 7.6) for 2 hours at RT while gently shaking. Slices were then stained with 10 μM DAPI (4',6-Diamidino-2-Phenylindole, Dihydrochloride, LifeTechnologies, CA, D1306) in 3 ml PBST$_{0.1}$. To compare the achievable imaging depth before and after clearing with 47% TDE/PBS, stacks of 600 μm depth with a *z* step of 2 μm were acquired with the TPFM. Then imaging depth was quantified by the decay of the image contrast value with depth in cleared and uncleared samples. The image contrast of each frame was calculated with equation (1).

$$(1)\ Contrast = \sqrt{\frac{\sum_i (c_i \times (i - \bar{I})^2)}{C - 1}}$$

Where $c_i$ is the pixel count for intensity level *i* in an image (with *i* ranging between gray levels 0 and 255). $\bar{I} = I/C$ is the average intensity of the image with *I* defined as the image intensity integral $I = \sum_i i \times c_i$ and *C* as the total pixel count $C = \sum_i c_i$.

**Transmission electron microscopy**

Using a vibratome (Vibratome 1000 Plus, Intracel LTD, UK), sections of 500 μm thickness were cut from a Thy1-GFP-M Glut-PFA-fixed mouse brain. Slices were incubated in PBS or in 47% TDE/PBS for 4 days at 37°C while gently shaking. Samples were washed with 50 ml of 0.01M PBS for 30 minutes and 3 times for 5 minutes each in 20 ml of 0.1 M cacodylate buffer (pH 7.4). Post-fixation, en-bloc staining and resin embedding for transmission electron microscopy sectioning and imaging were performed following Knott[28]. A fixation of 40 minutes with 1% osmium tetroxide in 0.1M cacodylate buffer (pH 7.4) was performed at RT. Sections were then washed twice for 5 minutes in distilled water followed by 10 minutes of 1% aqueous uranyl acetate. Sections were dehydrated in graded alcohol series (2x50%, 1x70%, 1x90%, 1x95%, 2x100%) for 3 minutes each with a final step of 10 minutes in propylene oxide. Sections were then embedded in EPON through an incubation of 1 hour in 1:1 propylene oxide:EPON, two incubations of 30 minutes in 100% EPON and one incubation of 4 hours in fresh 100% EPON. Finally, sections were placed in fresh EPON and incubated for 24 hours at 65°C to allow resin polymerization. Images were obtained with a TEM microscope (TEM CM 12, PHILIPS).

**Two-photon fluorescence microscopy**

A mode locked Ti:Sapphire laser (Chameleon, 120 fs pulse width, 90 MHz repetition rate, Coherent, CA) was coupled into a custom-made scanning system based on a pair of galvanometric mirrors (VM500+, Cambridge Technologies, MA). The laser was focused onto the specimen by a water immersion 20x objective lens (XLUM 20, NA 0.95, WD 2mm, Olympus, Japan) for uncleared (PBS) sample imaging or a tunable 20x objective lens (Sca*l*e LD SC Plan-Apochromat, NA 1, WD 5.6mm, Zeiss, Germany) for cleared (47% TDE/PBS) sample imaging. The system was equipped with a motorized *xy* stage (MPC-200, Shutter Instrumente, CA) for axial displacement of the sample and with a closed-loop piezoelectric stage (ND72Z2LAQ PIFOC objective scanning system, 2mm travel range, Physik Instrumente, Germany) for the displacement of the objective along the *z* axis. The fluorescence signals were collected by two photomultiplier modules (H7422, Hamamatsu Photonics, NJ). The instrument was controlled by custom software, written in LabView (National Instruments, TX).

**Serial two-photon tomography with TDE**

The hippocampus was manually dissected from a PFA-fixed, adult (p56), Thy1-GFP-M mouse brain and cleared by two serial incubations in 20 ml of 20% and 47% (vol/vol) TDE/PBS, each for 1 hour at 37°C while gently shaking. After clearing, the hippocampus was embedded in a solution of 47% TDE/PBS (vol/vol) – agar 4% (wt/vol). In order to reduce the number of slices required, the hippocampus was horizontally oriented with respect to the optical planes acquired. Serial sectioning was performed with a vibratome and after every cut the sample was left in 47% TDE/PBS overnight at RT. Stacks of each layer were acquired with the TPFM (Zeiss 20X Sca*l*e objective, pixels size 0.59 x 0.59 $\mu m^2$) using a custom LabView program (National Instruments) allowing for automatic acquisition of adjacent regions drawing a spiral square. Each stack had a depth of 1000 µm with a z displacement of 4 µm between images. Each frame had a field of view of 300 x 300 $\mu m^2$, adjacent stacks had an overlap of 30 µm. To ensure an efficient 3D reconstruction along the *z* axis, slicing was performed every 800 µm such that the subsequent layer had an overlapping region of 200 µm with the previous one. To obtain constant fluorescence intensity, laser power was increased during acquisition according to the imaging depth, however, some illumination inhomogeneity was present due to the inherent heterogeneity of the tissue.

**Light-sheet microscopy**

Specimens were imaged using a custom-made confocal light sheet microscope (CLSM) described in Silvestri[29]. The light sheet was generated by scanning the excitation beam with a galvanometric mirror (6220H, Cambridge Technology, MA) and confocality was achieved by synchronizing the galvo scanner with the line read-out of the sCMOS camera (Orca Flash4.0, Hamamatsu Photonics, Japan). Five different cw wavelengths were available (MLDs and DPSSs, Cobolt, Sweden) for fluorescence excitation and an acousto-optic tunable filter (AOTFnC-400.650-TN, AA Opto-Electronic, France) was used to regulate laser power. The excitation light was focused with a long working distance, low magnification objective (10x 0.3NA WD 17.5mm, Nikon, Japan) and fluorescence was collected on a perpendicular axis with a specialized objective for high refractive index immersion and a correction collar for refractive indices ranging from 1.41 to 1.52 (XLSLPLN25XGMP, 25x 1.0NA, WD 8mm Olympus, Japan). The samples were mounted on a motorized x-, y-, z-, θ-stage (M-122.2DD and M-116.DG, Physik Instrumente, Germany) which allowed free 3D motion plus rotation in a custom-made chamber filled with 63% TDE/PBS. The microscope was controlled via custom written LabVIEW code (National Instruments) which coordinated the galvo scanners, the rolling shutter and the stack acquisition.

**Image processing**

In order to achieve a 3D image of the whole specimen from raw data, the Terastitcher[30] has been recently proposed, i.e. a stitching tool capable to deal with teravoxel-sized images. However, the Terastitcher does not support input data acquired through the serial sectioning procedure, which leads to a specimen partitioned in different layers. Furthermore, only single channel images can be processed. For these reasons, we extended the Terastitcher functionalities introducing the two following additional features: i) stitching of a specimen partitioned in a number of overlapping layers for the hippocampus reconstruction and, ii) coping with images containing more than one channel for the human brain tomography. With respect to the first requirement, that is allowing a complete reconstruction of a multi-layered raw data, we schematically depict the adopted strategy in supplementary figure 4. First of all, the various input layers, each of which is composed of several parallel overlapping stacks, were separately stitched using the existing Terastitcher tool. After this preliminary step, leveraging the layer coordinates provided by the instrument, we imported the processed layers as a new volume where each layer had a partial overlap with adjacent layers. Furthermore, each layer was organized in a non-overlapping tiled format, enabling the application of a multi MIP-NCC approach[30], which computes the displacement between two adjacent layers along all of the three directions. When a displacement computation for each pair of adjacent layers has been computed, the overlapping regions were merged through a blending procedure that smooths the transition between layers, resembling once more the procedure used by Terastitcher for combining adjacent stacks. Finally, we noted that due to the repositioning of each layer, the volume containing the reconstructed specimen could contain empty regions, which were thereby filled with black voxels. As to the second additional feature, that is handling multi-channel images, the Terastitcher has been extended so that the MIP-NCC algorithm used for displacement computation could work on an image, which could be either the fusion of the input channels or one of the input channels. This permitted to privilege the channel with more information content over the other channels as well as discard noisy channels. The reconstructed 3D image can then be produced with the same channel composition of raw data.

**Data analysis**

Graphs and data analysis were done with OriginPro 9.0 (OriginLab Corporation). Stacks were analyzed using both Fiji (http://fiji.sc/Fiji) and Amira 5.3 (Visage Imaging) software. 3D renderings of stitched images were produced using the Amira Voltex function. The Filament Editor of Amira was used to manually trace neuronal filaments.

**REFERENCES**

1  Lein, E. S. *et al.* Genome-wide atlas of gene expression in the adult mouse brain. *Nature* **445**, 168-176 (2007).
2  Bohland, J. W. *et al.* A proposal for a coordinated effort for the determination of brainwide neuroanatomical connectivity in model organisms at a mesoscopic scale. *PLoS Comput Biol* **5**, e1000334 (2009).
3  Li, A. *et al.* Micro-optical sectioning tomography to obtain a high-resolution atlas of the mouse brain. *Science* **330**, 1404-1408 (2010).
4  Gong, H. *et al.* Continuously tracing brain-wide long-distance axonal projections in mice at a one-micron voxel resolution. *Neuroimage* **74**, 87-98 (2013).
5  Ragan, T. *et al.* Serial two-photon tomography for automated ex vivo mouse brain imaging. *Nat Methods* **9**, 255-258 (2012).
6  Huisken, J. & Stainier, D. Y. Selective plane illumination microscopy techniques in developmental biology. *Development* **136**, 1963-1975 (2009).
7  Dodt, H. U. *et al.* Ultramicroscopy: three-dimensional visualization of neuronal networks in the whole mouse brain. *Nat Methods* **4**, 331-336 (2007).
8  Becker, K., Jahrling, N., Saghafi, S., Weiler, R. & Dodt, H. U. Chemical clearing and dehydration of GFP expressing mouse brains. *PLoS One* **7**, e33916 (2012).
9  Renier, N. *et al.* iDISCO: A Simple, Rapid Method to Immunolabel Large Tissue Samples for Volume Imaging. *Cell* **159**, 896-910 (2014).
10 Hama, H. *et al.* Scale: a chemical approach for fluorescence imaging and reconstruction of transparent mouse brain. *Nat Neurosci* **14**, 1481-1488 (2011).
11 Ke, M. T., Fujimoto, S. & Imai, T. SeeDB: a simple and morphology-preserving optical clearing agent for neuronal circuit reconstruction. *Nat Neurosci* **16**, 1154-1161 (2013).
12 Kuwajima, T. *et al.* ClearT: a detergent- and solvent-free clearing method for neuronal and non-neuronal tissue. *Development* **140**, 1364-1368 (2013).
13 Susaki, E. A. *et al.* Whole-brain imaging with single-cell resolution using chemical cocktails and computational analysis. *Cell* **157**, 726-739 (2014).
14 Chung, K. *et al.* Structural and molecular interrogation of intact biological systems. *Nature* **497**, 332-337 (2013).
15 Tomer, R., Ye, L., Hsueh, B. & Deisseroth, K. Advanced CLARITY for rapid and high-resolution imaging of intact tissues. *Nat Protoc* **9**, 1682-1697 (2014).
16 Staudt, T., Lang, M. C., Medda, R., Engelhardt, J. & Hell, S. W. 2,2'-thiodiethanol: a new water soluble mounting medium for high resolution optical microscopy. *Microsc Res Tech* **70**, 1-9 (2007).
17 Appleton, P. L., Quyn, A. J., Swift, S. & Nathke, I. Preparation of wholemount mouse intestine for high-resolution three-dimensional imaging using two-photon microscopy. *J Microsc* **234**, 196-204 (2009).
18 Gonzalez-Bellido, P. T. & Wardill, T. J. Labeling and confocal imaging of neurons in thick invertebrate tissue samples. *Cold Spring Harb Protoc* **2012**, 969-983 (2012).
19 Frasconi, P. *et al.* Large-scale automated identification of mouse brain cells in confocal light sheet microscopy images. *Bioinformatics* **30**, i587-593 (2014).
20 Baek, S. T., Gibbs, E. M., Gleeson, J. G. & Mathern, G. W. Hemimegalencephaly, a paradigm for somatic postzygotic neurodevelopmental disorders. *Current opinion in neurology* **26**, 122-127 (2013).
21 Poguzhelskaya, E., Artamonov, D., Bolshakova, A., Vlasova, O. & Bezprozvanny, I. Simplified method to perform CLARITY imaging. *Molecular neurodegeneration* **9**, 19 (2014).
22 Silvestri, L., Allegra Mascaro, A. L., Costantini, I., Sacconi, L. & Pavone, F. S. Correlative two-photon and light sheet microscopy. *Methods* **66**, 268-272 (2014).
23 Jahrling, N., Becker, K., Schonbauer, C., Schnorrer, F. & Dodt, H. U. Three-dimensional reconstruction and segmentation of intact Drosophila by ultramicroscopy. *Front Syst Neurosci* **4**, 1 (2010).

24   Huisken, J., Swoger, J., Del Bene, F., Wittbrodt, J. & Stelzer, E. H. Optical sectioning deep inside live embryos by selective plane illumination microscopy. *Science* **305**, 1007-1009 (2004).
25   Feng, G. *et al.* Imaging neuronal subsets in transgenic mice expressing multiple spectral variants of GFP. *Neuron* **28**, 41-51 (2000).
26   Taniguchi, H. *et al.* A resource of Cre driver lines for genetic targeting of GABAergic neurons in cerebral cortex. *Neuron* **71**, 995-1013 (2011).
27   Madisen, L. *et al.* A robust and high-throughput Cre reporting and characterization system for the whole mouse brain. *Nat Neurosci* **13**, 133-140 (2010).
28   Knott, G. W., Holtmaat, A., Trachtenberg, J. T., Svoboda, K. & Welker, E. A protocol for preparing GFP-labeled neurons previously imaged in vivo and in slice preparations for light and electron microscopic analysis. *Nat Protoc* **4**, 1145-1156 (2009).
29   Silvestri, L., Bria, A., Sacconi, L., Iannello, G. & Pavone, F. S. Confocal light sheet microscopy: micron-scale neuroanatomy of the entire mouse brain. *Opt Express* **20**, 20582-20598 (2012).
30   Bria, A. & Iannello, G. TeraStitcher - a tool for fast automatic 3D-stitching of teravoxel-sized microscopy images. *BMC Bioinformatics* **13**, 316 (2012).

**ACKNOWLEDGEMENTS**

The research leading to these results has received funding from the European Union Seventh Framework Program (FP7/2007-2013) under grant agreements no. 604102 (Human Brain Project) and n° 284464 (LASERLAB-EUROPE). The research has also been supported by the Italian Ministry for Education, University and Research in the framework of the Flagship Project NANOMAX, by ''Ente Cassa di Risparmio di Firenze'' (private foundation) and by Regione Toscana (grant number: POR-CreO 2007–2013). We thank Prof. Giovanni Delfino for discussions on the electron microscopy data and Marcel van 't Hoff for his help on LabVIEW code programming.

**AUTHOR CONTRIBUTION**

I.C. and L.Sac. planned the experiments; I.C., J.P.G. and A.P.D.G. prepared the clarified samples; I.C. and A.P.D.G. performed the characterization of the agent; I.C., A.L.A.M. imaged the samples with TPFM; J.P.G., L.Sil., M.C.M. imaged the samples with LSM; L.O. and G.I. processed the images; F.V. performed the TEM imaging; V.C. prepared the human brain sample; G.I., R.G., H.M. and F.S.P. supervised the project; I.C. made the figures and wrote the paper with input from all other authors.

**COMPETING FINANCIAL INTERESTS**

The authors declare no competing financial interests.

**FIGURE LEGENDS**

**Figure 1. TDE characterization.** (a) Transmission images of 1 mm thick hemi-brain slices of Thy1-GFP-M mouse in PBS and after clearing with various solutions. (b) Light transmittance curves of hemi-brain slices shown in a (mean ± s.e.m., n=4). (c) Normalized linear deformation during optical clearing (mean ± s.d., n=4). (d) Contrast decay as function of depth in uncleared and cleared samples (mean ± s.d., n=10). (e) Half-time fluorescence decay (mean ± s.d., n=4); clearing did not increase photobleaching compared to PBS. (f) Fluorescence intensity over time (mean ± s.d., n=10); no quenching effect was observed after incubation of the sample in 47% TDE/PBS for up to two months. (g) Two-photon fluorescence imaging of 2 mm FVB mouse brain slices stained with DAPI in PBS and in 47% TDE/PBS. Reconstruction along z axis, depth 1 mm; scale bar = 100 μm. (h) TEM images of Thy1-GFP-M mouse brain slices previously incubated in PBS and in 47% TDE/PBS for 4 days. Triangles indicate mitochondria (red), axons (green) and nuclei (blue). Scale bar = 2 μm (upper panels) and 200 nm (lower panels).

**Figure 2. Hippocampus tomography.** Reconstruction of entire Thy1-GFP-M mouse hippocampus fixed with PFA and cleared with 47% TDE/PBS (Zeiss 20X Sca*le* objective, two-photon excitation). (a) 3D rendering of six layers of 1 mm thickness each, sampled every 4 μm. Serial sectioning at 800 μm depth. Scale of figure can be inferred from the white cube in bottom right corner which has 300 μm side. (b) Coronal section corresponding to the yellow box in a. It is possible to recognize specific anatomical features of the hippocampus; DG: dentate gyrus, CA1 and CA3: *Cornu Ammonis* areas; scale bar = 300 μm. (c) Single image at full resolution of one stack, demonstrating the visualization of fine dendritic and axonal fibers; scale bar = 50 μm. (d) and (e) Magnified insets corresponding to red boxes in c. Red triangles highlight axon varicosities and dendritic spines; scale bar = 10 μm. (f,g) Three-dimensional tracing of single neurons through different stacks and layers of the hippocampus. Image in f shows maximum intensity projection of volume between two planes highlighted in a with the red rectangles; scale bar = 300 μm.

**Figure 3. TDE refractive index matching for CLARITY.** (a) Transmission images of a Thy1-GFP-M mouse brain after CLARITY protocol: hydrogel–tissue hybridization and lipid removal with ETC, cleared with various solutions. (b) Light transmittance of 2 mm CLARITY brain slices in different solutions (mean ± s.e.m., n=4). Transmittance increased with the RI and that of 63% TDE/PBS is comparable with that of FocusClear[TM]. (c) Normalized linear deformation of CLARITY brains in different solutions (mean ± s.d., n=4). After initial expansion due to the ETC lipid removal, the tissue shrank back to its original size during the clearing step.

**Figure 4. Whole mouse brain tomography.** Imaging of whole transgenic mouse brains treated with CLARITY and cleared with 63% TDE/PBS imaged with LSM (Olympus, 25X objective). (a) 3D rendering of a parvalbumin- tdTomato brain. (b) 3D rendering of stacks from PV-tdTomato mouse brain, GAD-tdTomato mouse brain, PI stained mouse brain, FITC-albumin labeled mouse brain, scale bar = 400 µm. (c,d,e,f) High resolution insert of stack corresponding to red boxes in c. Scale bar = 100 µm.

**Figure 5. Human brain immunostaining.** 2 mm thick block of formalin-fixed tissue removed from the dysplastic hemisphere of a patient with hemimegalencephaly, treated with PC CLARITY protocol, immunostained with different antibodies and cleared with 47% TDE/PBS (Zeiss 20X Sca*l*e objective, two-photon excitation). (a) Parvalbumin (PV) staining in red and nuclei (DAPI) in cyan; scale bar = 100 µm. (b) Glial fibrillary acidic protein (GFAP) staining in yellow and nuclei (DAPI) in cyan; scale bar = 100 µm. (c) GFAP staining in yellow and PV in red; scale bar = 100 µm. (d,e) 3D rendering of a 1 x 1 x 1 mm$^3$ and 0.25 x 0.25 x 1 mm$^3$ section of brain tissue labeled for PV (red) and DAPI (cyan). (f) Horizontal view of sections of e at different depths; scale bar = 100 µm. (g,h) Three-dimensional tracing of parvalbumin fibers through the volume shown in d. Scale bar = 200 µm.

**TABLE**

| Method | Clearing composition | Refractive index | Linear deformation | Protein fluorescence quenching | Sample characteristic | High viscosity | Cost* | Clearing time | Immunostaining compatibility | Clearing capability | References |
|---|---|---|---|---|---|---|---|---|---|---|---|
| Et + BABB | Ethanol Benzyl alcohol Benzyl Benzoate | 1.54 | Shrinkage | Yes | Stiff | No | $ | 2 days | No | Good | 7 |
| THF + DBE | Thetrahydrofuran Dibenzyl ether | 1.56 | Shrinkage | Yes | Stiff | No | $ | 2 days | No | Good | 8 |
| iDISCO | Thetrahydrofuran Dibenzyl ether | 1.56 | Shrinkage | Yes | Stiff | No | $$ | 2 days | Yes | Good | 9 |
| Glycerol | Glycerol solution | 1.44 | No | No | / | Yes | $ | 2 days | No | Moderate | / |
| Sca/e | Urea Glycerol Triton X-100 | 1.38 | Expansion | No | Fragile | No | $ | Months | No | Moderate | 10 |
| SeeDB | Fructose saturated α thioglycerol | 1.49 | No | No | / | Yes | $ | 2 weeks | No | Moderate | 11 |
| ClearT | Formamide Polyethylene glycol | 1.45 | No | No | / | No | $ | 1 day | No | Moderate | 12 |
| CUBIC | Urea Aminoalcohols | 1.47 | Transient swelling | No | / | No | $$ | 2 weeks | Yes | Good | 13 |
| CLARITY | FocusClear | 1.45 | Transient swelling | No | Spongy | No | $$$ | 10 days | Yes | Good | 14 |
| CLARITY2 | PBST | 1.33 | / | No | Spongy | No | $$ | 12 days | Yes | Good | 21 |
| TDE | 2,2' Thiodioethanol | Tunable | No | No | / | No | $ | 2 hours | No | Moderate | / |
| CARITY + TDE | 2,2' Thiodioethanol | Tunable | Transient swelling | No | Spongy | No | $$ | 10 days | Yes | Good | / |

**Table 1. Comparison of recently published clearing methods.** Main characteristics of different clearing methods reported in literature. We considered adult mouse brain as typical sample. This table gives an overview of available techniques and their limitations and advantages. *Cost was considered for both



labeling and clearing and was indicated as inexpensive ($) if less than $50, medium ($$) if between $50 and $500, and expensive ($$$) for over $500 per sample.

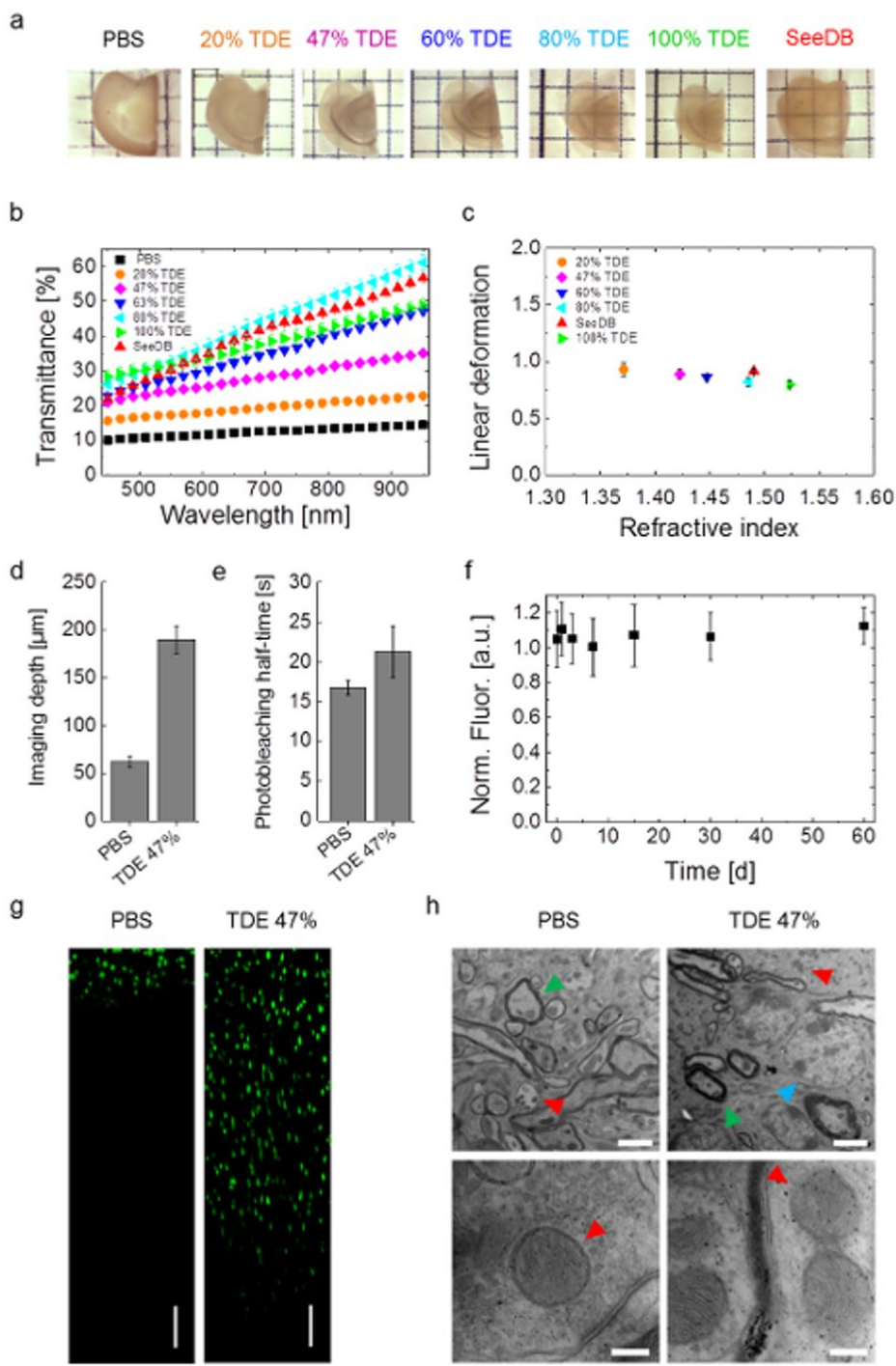

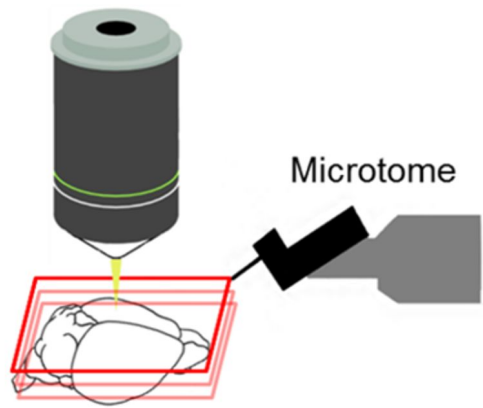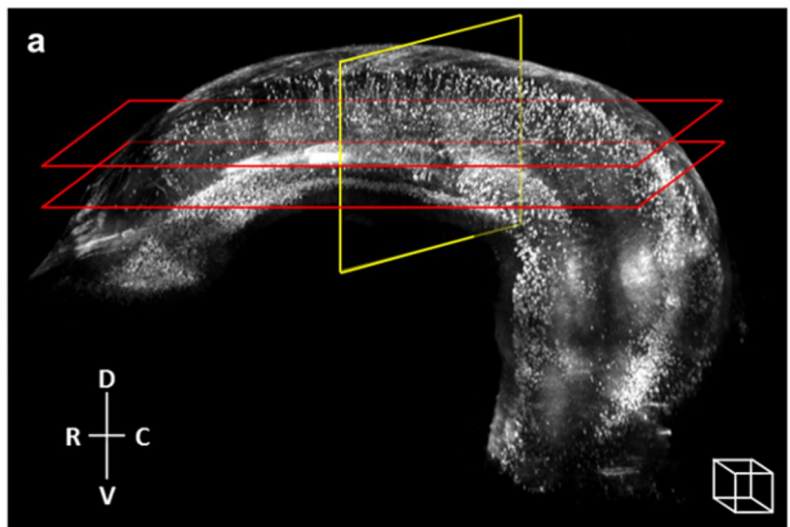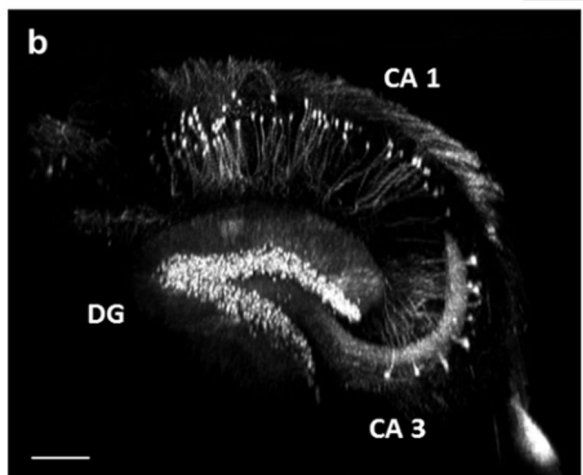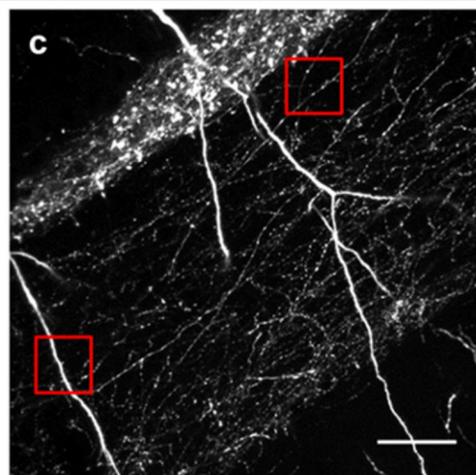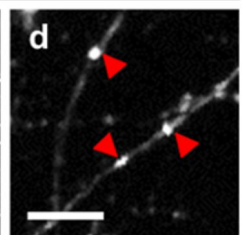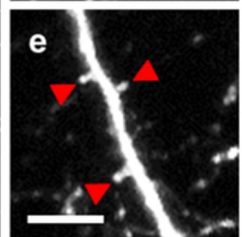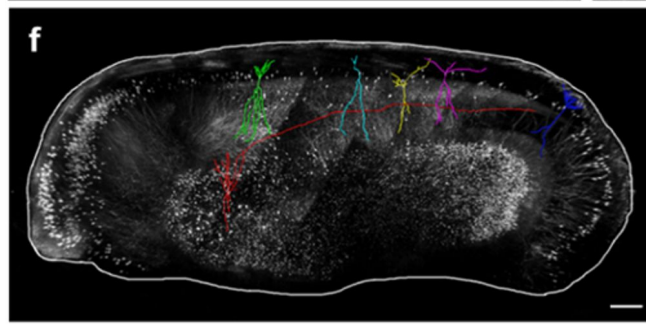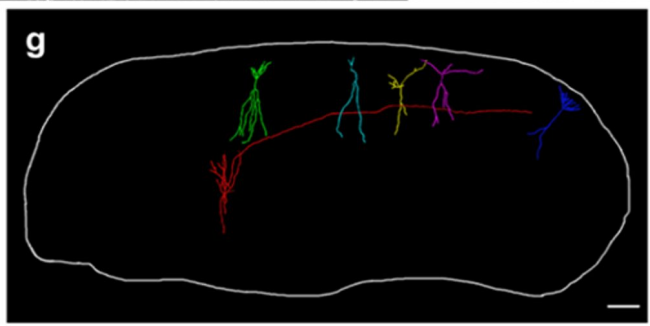

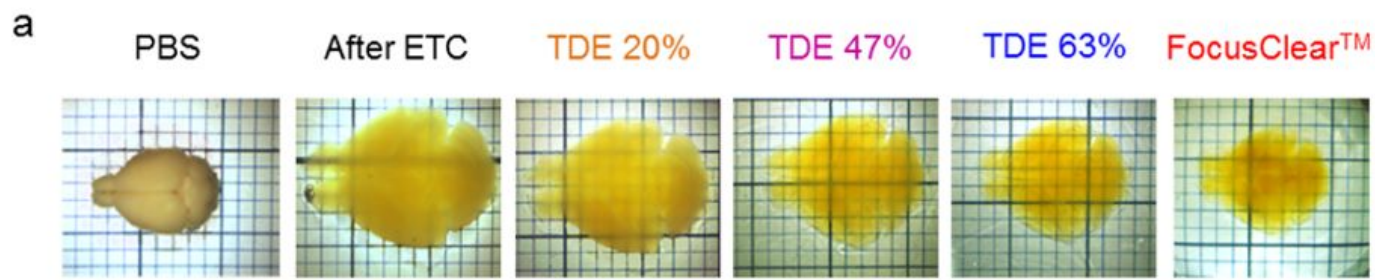

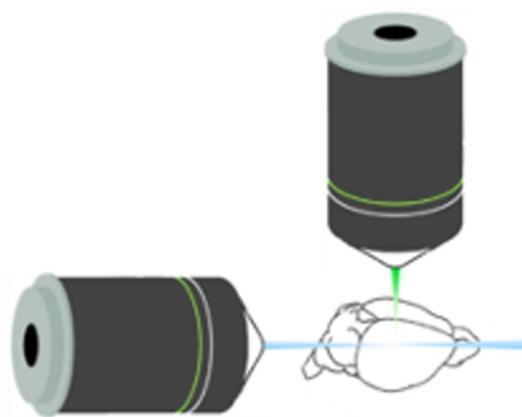
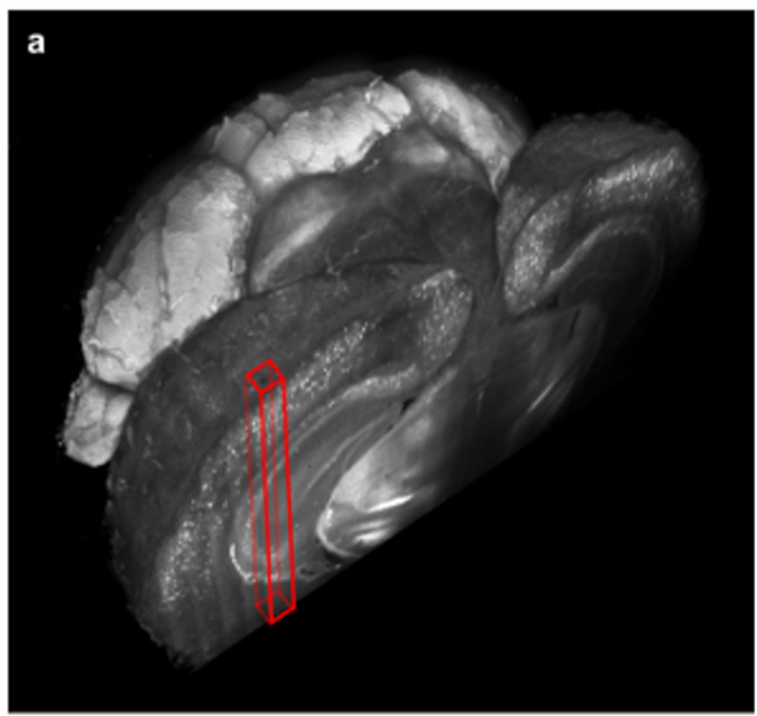
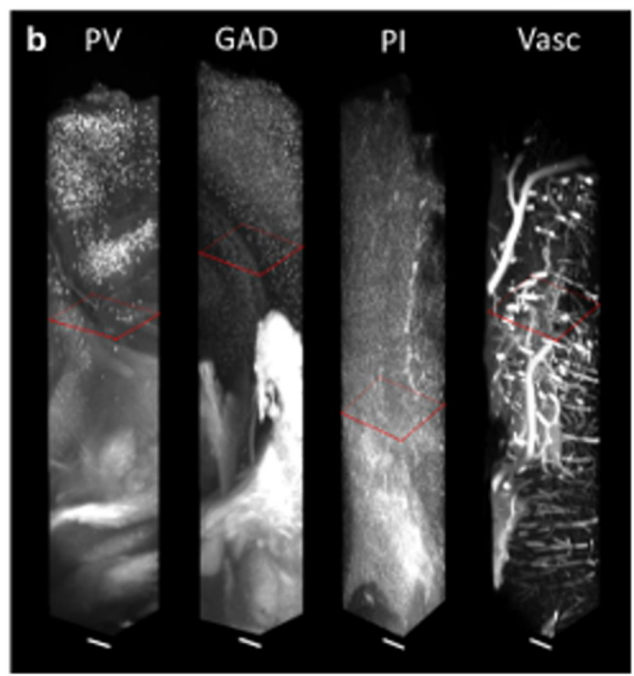
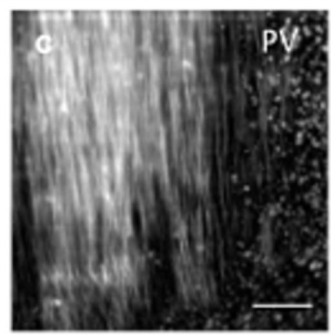
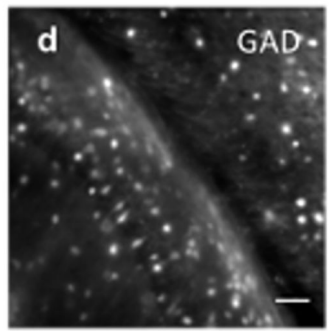
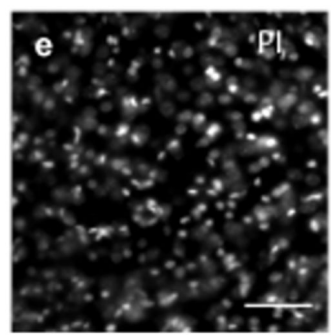
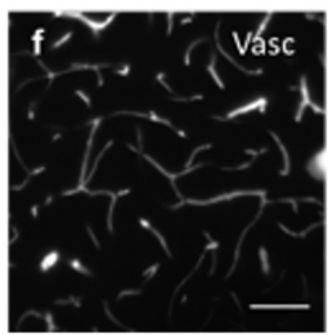

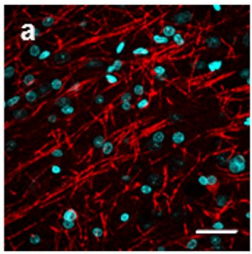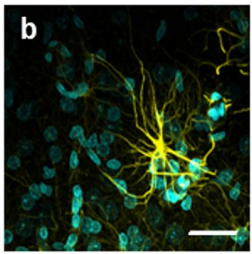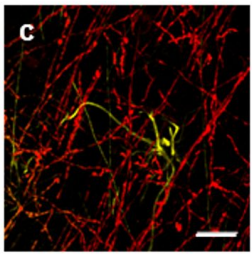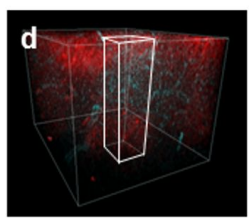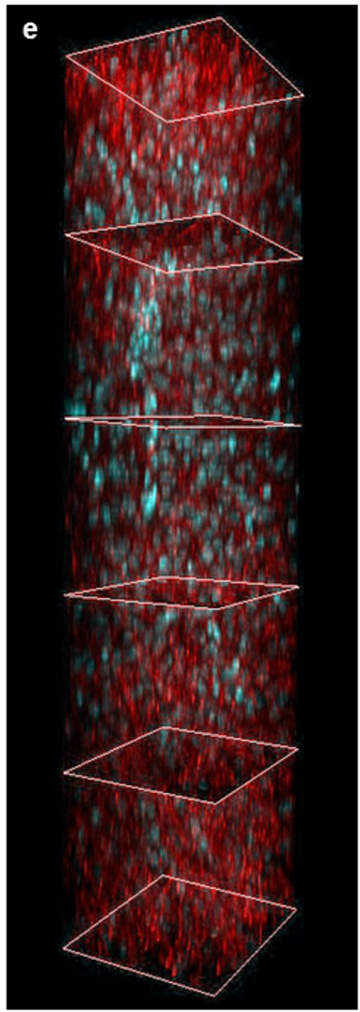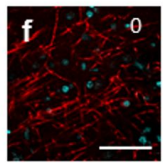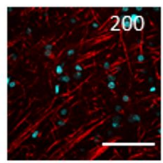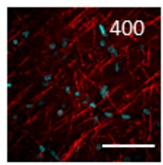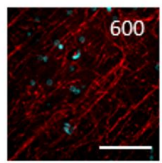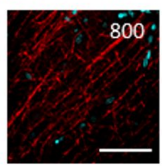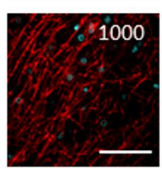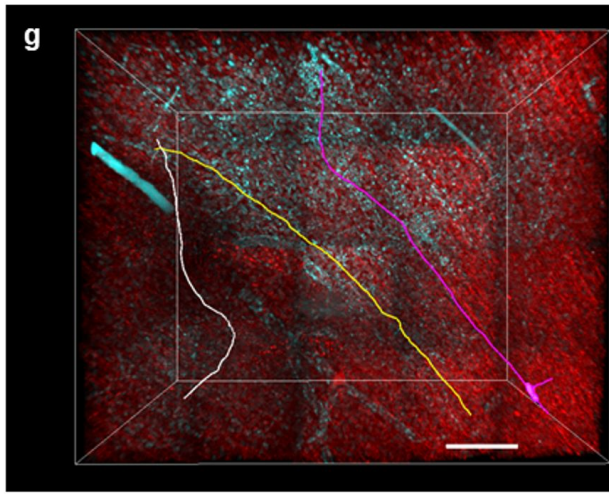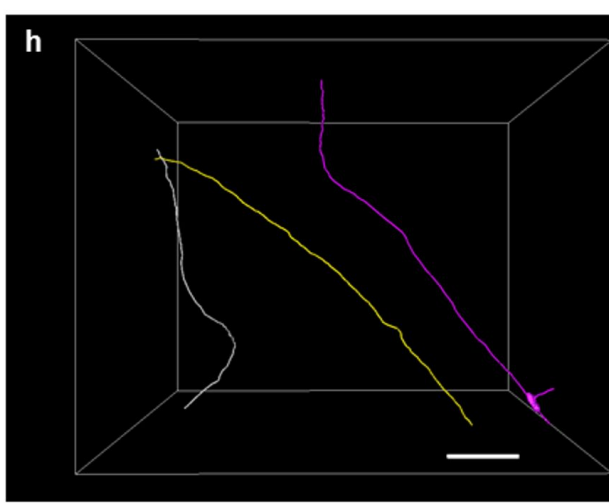